# Adaptive Traffic Signal Control based on Multi-Agent Reinforcement Learning. Case Study on a Simulated Real-World Corridor

Dickens Kakitahi Kwesiga[1], Angshuman Guin[1], Michael Hunter[1]

*Abstract*—**Previous studies that have formulated reinforcement learning (RL) algorithms for adaptive traffic signal control have mainly been limited to single intersection control and very simplified signal timing plans. RL approaches remain untested for the more complex, coordinated and realistic field signal control scenarios. The current study attempts to address these gaps and formulates a multi-agent proximal policy optimization (MA-PPO) algorithm to implement adaptive and coordinated traffic control along an arterial corridor. The formulated MA-PPO has a centralized-critic architecture under a centralized training and decentralized execution framework. Agents are designed to allow selection and implementation of up to eight signal phases, as commonly implemented in field controllers. The formulated algorithm is tested on a simulated real-world seven intersection corridor. The speed of convergence for each agent was found to depend on the size of the action space, which depends on the number and sequence of signal phases. The performance of the formulated MA-PPO adaptive control algorithm is compared with the field implemented actuated-coordinated signal control (ASC), modeled using PTV-Vissim®-MaxTime® software in the loop simulation (SILs). The trained MA-PPO performed significantly better than the ASC for all movements. Compared to ASC the MA-PPO showed 2% and 24% improvements in travel time in the primary and secondary coordination directions, respectively. For cross streets movements MA-PPO also showed significant crossing time reductions. Volume sensitivity experiments revealed that the formulated MA-PPO demonstrated good stability, robustness, and adaptability to changes in traffic demand.**

Dickens Kakitahi Kwesiga, Angshuman Guin and Michael Hunter are with the school of Civil and Environmental Engineering, Georgia Institute of Technology, Atlanta GA, 30332 USA (email: dkwesiga3@gatech.edu; angshuman.guin@ce.gatech.edu; michael.hunter@ce.gatech.edu)

## I. INTRODUCTION

With the rise of complex and large machine learning (ML) models, there is heightened interest in developing ML based adaptive traffic signal control systems. Compared to other ML approaches, reinforcement learning (RL) is a direct choice for control problems such traffic signal control. Once trained, model free RL requires less computational resources to implement in real time compared to existing adaptive signal control systems based on mathematical programming.

Previous efforts to develop adaptive RL-based signal control systems have mainly been limited to single agent / single intersection formulations with only a few recent efforts extending the formulation to multiple agents /multiple intersections. Additionally, previous studies have almost universally considered very simplified signal timing plans including limiting the number of phases, fixing the phasing sequences and fixing the duration of each signal phase. Therefore, for most part, RL approaches remain untested for the more complex, coordinated and realistic field signal control scenarios. The current study formulates multi-agent reinforcement learning (MARL) based adaptive signal algorithms considering real-world signal timing and network complexities. The formulated algorithms are tested on a real-world simulated arterial corridor and their performance compared to the field deployed optimal actuated-coordinated signal control (ASC) plans.

### A. Related Studies

RL is an approach of machine learning (ML) in which the learning entity, also called the agent, learns by continuous interaction with the environment. RL seeks to find the optimal mapping of states to actions, to achieve the highest numerical rewards. By trying out different actions over time the agent learns to identify the actions that maximize rewards given the state of the environment. MARL involves multiple agents interacting with a common environment and with each other. Moving from single agent to MARL introduces new challenges that require modification or formulation of new learning frameworks. These challenges include equilibrium selection, non-stationarity of the environment, scaling to many agents, multi-agent credit assignment, and partial observability [1], [2].

Several studies have formulated and tested RL-based adaptive traffic signal control algorithms in simulation environments. These studies show that RL-based algorithms can potentially outperform conventional fixed and actuated signal timing [3], [4], [5], [6], [7], [8], [9], [10], [11], [12]. A few recent efforts have attempted to extend RL-based traffic control from a single intersection to multiple intersections and networks [8], [13], [14], [15], [16], [17].

Most of the reviewed studies assume simplified signal plans including limiting the number of allowable phases [3], [4], [5], [8], [9], [10], [12], [13], [16], [18], assuming fixed timing sequences [3], [4], [5], [6], [7], [8], [9], [10], [12], [13], [14], [16], [18], [19], [20], fixing the duration or intervals of green for each selected phase or phase combination [4], [8], [14], [21], and utilizing single ring barrier configurations [4], [8], [14], [21], [22], [23], [24]. Therefore, for most part, RL approaches remain untested for the more complex and realistic field signal control scenarios.

The biggest share of reviewed studies formulate and trains RL control agents based on deep q-networks (DQN) and its variations [5], [8], [9], [11], [16], [18], [21], [23], [25], [26], [27], [28]. The reviewed MARL studies are almost universally based on DQN [13], [14], [15], [17], [19], [27], [29]. However, recent literature shows that policy gradient based methods show better performance and convergence, especially in partially observable environments [30], [31]. In a study intriguingly entitled "the surprising effectiveness of PPO in cooperative multi-agent games", the authors compared the performance of multi-agent proximal policy optimization (PPO) and other centralized training decentralized execution (CTDE) algorithms, including Q-mix, for four multi-agent environments [31]. PPO showed comparable or superior performance and required minimum hyperparameter tuning compared to other algorithms.

Signal coordination is at the heart of arterial traffic flow optimization. A key objective of MARL-based traffic control algorithms is to promote cooperation between individual intersection agents to progress traffic through adjacent intersections. There is a very limited number of studies for MARL-based traffic control. Of the few studies, some have developed q-mix based algorithms [13], [14], [15], [19]. decomposition (VDN) [27], [32] while others have formulated independent q-learning algorithms [17], [29]. However, as indicated above, policy gradient (PG) algorithms, such as MA-PPO, are likely to achieve better performance in partially observable multi-agent environments compared to value-based methods [1], [30].

### *B. Problem Definition and Objectives*

The current study formulates a multi-agent proximal policy optimization (MA-PPO) adaptive signal control algorithm and tests the algorithm on a simulated real-world corridor with field traffic volumes and geometry. The study seeks to overcome limitations identified in the literature, such as: (a) most studies are limited to single intersection control, (b) the few multiple intersections studies have almost universally formulated value-based RL algorithms, and (c) most studies utilize highly simplified signal control. Reinforcement learning approaches remain largely untested for more complex and normally coordinated real-world signal timing plans.

The formulated MA-PPO has a centralized critic architecture in which each agent is formulated with an actor network that selects independent actions conditioning on local observations and a centralized critic that estimates the agent's value function conditioning on global observations. All agents are formulated to allow the selection and implementation of up to eight signal phases, as commonly implemented in field controllers. For the test corridor and field measured traffic volumes, the performance of the formulated MA-PPO is compared against field implemented actuated-coordinated signal timings modeled using software in the loop simulation (SILs). To test the robustness of the formulated algorithm, sensitivity experiments are performed with volumes adjusted (a) upwards by 5% and (b) downwards by 10% from the field measured volumes

## II. METHODOLOGY

### *A. Proximal Policy Optimization (PPO) Overview*

#### *1. Single Agent PPO*

Central to much of RL theory and most RL algorithms are value functions. Value functions are used to define the expected returns of individual states and actions. The state value function, $V^{\pi}(s)$, gives the expected returns starting from state ($s$), and following the policy ($\pi$) thereafter. The action value function, $Q^{\pi}(s,a)$, gives the expected returns starting from state ($s$), taking action ($a$), and following the policy thereafter. Thus, the state value function is the expected value across all action value functions, for all possible actions under a given policy.

RL algorithms fall under broad categories of value-based methods and policy-based methods. Value-based methods learn value functions and to select actions, the agent follows a policy derived from the learned value function. In deep RL, value-based methods such as DQN learn parameterized value-functions using neural networks. Policy-based based methods directly learn parametrized policy functions. Policy-based methods can easily learn stochastic policies which comes with several advantages, including (a) better performance under partially observable environments, and (b) better convergence properties as the action probabilities change smoothly as a function of learned parameters because policies are parameterized with continuous values [30]. PG algorithms are policy-based algorithms in which gradient-based optimization is used to optimize the policy parameters.

PPO belongs to a family of PG algorithms called actor-critic. Actor-critic algorithms have an actor network that learns a parameterized policy function to select actions and a critic network that learns a parameterized value function to evaluate the actions selected by the actor network. For discrete action spaces, on the forward pass, the actor network takes the environment state as input and outputs a probability distribution over the discrete action set. During training PG methods compute gradients of the loss function and optimize the policy parameters to follow the direction of higher expected returns. Returns are evaluated by value functions, $V^{\pi}(s)$ and $q^{\pi}(s,a)$. The foundational actor-critic algorithm called advantage actor critic (A2C) computes estimates of advantage, *Adv(s,a)* using

$$Adv(s,a) = Q(s,a) - V(s) = \begin{cases} r^t & if\ s^{t+1}\ is\ terminal \\ r^t + \gamma V(s^{t+1}) - V(s^t) & otherwise \end{cases} \quad (1)$$

to guide policy gradients. In (1), $r^t$ is the reward at time step, *t* with all other parameters already defined. Equations (1) – (4), and (7) are adapted from [1] while (5) and (6) are adapted from[33]. In both cases the notation is modified.

Gradients are computed from the gradient theorem as in

$$\nabla_{\varphi} J(\varphi) = E_{s\sim \Pr(.|\pi), a\sim\pi(.|s;\varphi)}\left[Q^{\pi}(s,a)\frac{\nabla_{\varphi}\pi(a|s;\varphi)}{\pi(a|s;\varphi)}\right] \quad (2)$$

where J is the function measuring the quality of policy, *Pr(.|π)* represents the state visitation probability under the policy *π, s* is the state, and *φ* are the policy parameters [1].

The actor loss, *L(φ)* is computed as in

$$\mathrm{L}(\varphi) = -Adv(s^t,a^t)log\pi(a^t|s^t;\varphi) \quad (3)$$

with all terms defined previously.

On a forward pass, the critic takes the state as input and outputs a single value of the state value function. The critic optimization loss *L(ϑ)* is the squared error of the value estimate and the bootstrapped target estimate $y^t$ as in

$$L(\vartheta) = (y^t - V(s^t;\vartheta))$$
$$y^t = \begin{cases} r^t & if\ s^{t+1}\ is\ terminal \\ r^t + \gamma V^{\pi}(s^{t+1};\vartheta) & otherwise \end{cases} \quad (4)$$

where *ϑ* are the parameters of the critic network.

A potential challenge of PG methods is a significant change in policy from any individual gradient update which can result in instability. Trust region policy Optimization (TRPO) formulated by [34] and PPO formulated by [33] try to mitigate this risk. PPO builds on TRPO to formulate a computationally efficient surrogate objective function based on importance sampling weights, ρ(s,a) which is defined as the fraction of probabilities of selecting an action (a) in state (s) for the new and old policies as in.

$$\rho(s,a) = \frac{\pi_{new}(a|s;\varphi)}{\pi_{old}(a|s)} \quad (5)$$

The loss function *L(φ)* is computed by clipping ρ(s,a) to restrict big changes to the policy for a single gradient update as in

$$\mathcal{L}(\varphi) = -\min\left(\begin{matrix} \rho(s^t,a^t)Adv(s^t,a^t) \\ clip(\rho(s^t,a^t), 1-\epsilon, 1+\epsilon)Adv(s^t,a^t) \end{matrix}\right) \quad (6)$$

In (6), the hyperparameter ϵ is selected to restrict how much the policy is allowed to change in a single update. Using ρ(s,a), PPO is able to restrict the magnitude of policy updates and update the policy multiple times using the same data

*2. Multi-Agent PPO – Centralized Critic*

PPO with centralized critic belongs to the CTDE paradigm of MARL. Each agent's actor network selects local actions conditioning on only local observations or local observation histories. The loss function of the actor remains as described earlier. The critic network is only utilized during training and can therefore utilize information that may not be accessible or available during execution. This information can include the full state of the environment and any external data. Fig. 1, adapted from [1], shows the architecture of centralized critic for agent i conditioning on local observations of agent i ($s_i$) and on observations of all other agents in the environment ($s_{n-i}$) while still approximating the value function of agent i's policy. In this centralized state value-based critic architecture, each agent i gets a local reward $r_i$ and does not take actions from other agents as inputs, as in centralized action value-based critic architectures. Despite not taking action inputs, centralized state value-based critics can still approximate advantage functions that provide preference over particular functions [1] and come with much simplicity compared to centralized action value-based critic architectures.

The loss function for agent i's critic is as in

$$\mathcal{L}(\vartheta_i) = (y_i - V(s_i^t, s_{n-i}^t;\vartheta_i))^2$$
$$y_i = r_i^t + \gamma V(s_i^t, s_{n-i}^t;\vartheta_i) \quad (7)$$

where $\vartheta_i$ are the network parameters, and r is the agent local reward with other variables as defined before.

*B. Definition of State, Action, and Reward Functions*

*1. State Definition*

The problem is formulated as a decentralized partially observable Markov decision process (DecPoMDP). At each intersection, the intersection actor network has access to only local observations. The local state has two component vectors, (1) vehicle state ($Veh_{state}$) and signal state ($Sig_{state}$). The vehicle state $Veh_{state}$ is a vector of the number of vehicles (v) in each lane (n) on each approach (m) as in

$$Veh_{state} = \begin{bmatrix} v_{11} \\ v_{12} \\ . \\ . \\ v_{mn} \end{bmatrix}. \quad (8)$$

The study assumes the availability of connected vehicle (CV) data or video camera data at the intersection to deduce the total count of vehicles at all the approaches of the intersection. The considered approach distance is the distance to the upstream intersection.

The $Sig_{state}$ has entries corresponding to the total number of intersection signal phases ($n_p$) as in

$$Sig_{state} = \begin{bmatrix} 10 \\ 0 \\ 12 \\ . \\ 0 \end{bmatrix}. \quad (9)$$

The entry for phase/s receiving green time are populated with the duration of green that has elapsed while all other entries are populated with zero. For all intersections of the corridor, the study assumes the availability of real time SPaT data which is increasingly ubiquitous.

The state for the centralized critic at intersection *i* ($Sc_i$) is defined as

$$Sc_i = [\,[Veh_{state_i} + Sig_{state_i}] + [Veh_{state_1} + Sig_{state_1} + \cdots + [Veh_{state_n} + Sig_{state_n}]]\,. \quad (10)$$

The gloabal state in (10) consists of (1) the local state of the agent which consists of $Veh_{state}$ and $Sig_{state}$ as defined above and (2) a concatenation of local states at all *n* other intersections. The local state at the subject intersection *i* is placed in the first position. It is perhaps worth reiterating that the critic network is only used during training and not during execution. During execution the trained algorithm does not access the global state defined in (10). This allows for flexibility in the information that can be included in the global state, including external data, as this data will not be required during execution.

*2. Action Definition and Implementation*

In this study, at each time *t*, an action is defined as whether the signal keeps the current phase or switches to any other phase in the set of valid phases. When the action changes from remaining in the same phase in time t-1 to selecting a different phase in t, the current phase yellow and red clearance intervals are displayed followed by the minimum green for the next selected phase. Every phase will have set minimum green that must be served, the phase green may be longer but not shorter than the minimum. Yellow and red clearance intervals allow safe termination of the current phase.

As indicated earlier, most of the reviewed studies assume very simplified signal plans. To more realistically replicate the field implemented signal plans, this study defines agent actions based on full dual ring barrier control, with the eight phase configurations as shown in Fig. 2. In dual ring control one phase from each ring is active simultaneously, where Ring 1 consists of phases {1,2,3,4} and Ring 2 consists of phases {5,6,7,8}. Active phases must not conflict, thus the barriers are defined by the ends of phases [2,6] and [4,8]. When crossing a barrier both rings must change phases simultaneously. Accordingly, there are eight possible discrete actions {0,1,2,3,4,5,6,7} corresponding to eight paired phase combinations, {1,5},{1,6}, {2,5}, {2,6}, {3,7}, {3,8}, {4,7}, and {4,8}. Each action represents a compatible phase combination with one phase from each ring. For instance, action 0 calls for phase 1 in ring 1 and phase 5 in ring 2.

[23] describes action selection and implementation with a single ring barrier configuration. Action implementation for the dual ring barrier control in the current study is more complex as allowable actions may be limited at various points throughout the cycle or a phase as shown in Fig. 3. To implement dual ring control the terminology “active phase” to refer to the current phase and “committed phase” to refer to the next selected phase is adopted for each ring. The benefit of tracking active and committed phase is that the only situation these differ is when a phase is in yellow or red clear. Thus, yellow and red clear may be timed without specifically implementing these indications as separate actions. Rather than this approach most recent studies include yellow and red clearance as separate actions in the action set [22], [27], [32]. However, this approach increases in the number of actions, potentially lengthening the training period.

Additionally invalid action masking (IAM) is used to force an agent to select the current active phase when required by signal control constraints. For example, when a phase is in it minimum green, yellow, or clearance intervals, the action must be to maintain the current active phase. IAM is also utilized to implement other constraints, such as implementing the ring barrier constraints. IAM is implemented by replacing logits corresponding to invalid actions with large negative numbers to preclude the selection of these actions.

*3. Time Step Definition*

In implementing RL actions in a simulation environment, there are two time steps being implemented, (1) the simulation time step and (2) the RL algorithm time step. The simulation time step defines the frequency of updating the simulation environment. In the simulation the time step is set to 0.1 second as this provides reasonable flow models and is the fidelity of field signal timing. The MARL algorithm time step defines the frequency of actions by the MARL agents. The size of the MARL algorithm time step ($\Delta t$) is not constant as defined next. Unless otherwise stated, in this paper, the term time step or variable $\Delta t$ refers to the MARL algorithm time step. The size of $\Delta t$ determines the frequency of interaction with the simulation which largely determines the run time and the training duration [23]. It is thus critical to optimize $\Delta t$ to improve run time efficiency.

To determine the MARL time step $\Delta t$, the time step for each agent, $\Delta ta_i$, must be determined, where there are 1 to i agents in the MARL. To determine $\Delta ta_i$ for an agent, it is necessary to first determine the allowable time step for each agent ring $\Delta ta_i r_j$. where there are j rings in agent i. Next, $\Delta ta_i r_j$ for ring j depends on the state and duration of active phase and the committed phase in that ring. If the active phase is serving the clearance time, $\Delta ta_i r_j$ for ring j is equal to the remaining clearance time plus the minimum green of the committed phase. If the active phase is running minimum green, $\Delta ta_i r_j$ for ring j becomes the remaining minimum green duration. If the active green is running green extension, $\Delta ta_i r_j$ is set to 1 second. The agent $\Delta ta_i$ is determined as the minimum of the two ring time steps, i.e., min ($\Delta ta_i r_1$ , $\Delta ta_i r_2$). See example illustrations in Fig. 3. Considering all agents in the environment, the overall $\Delta t$ is selected as the minimum $\Delta ta_i$ for all agents, i.e., minimum ($\Delta ta_1, \Delta ta_2, \ldots \Delta ta_i$). The more agents there are in the environment the likely smaller the $\Delta t$ and consequently the larger the run time. The minimum allowable $\Delta t$ is 0.1 second which corresponds to the simulation resolution of 10 steps per second.

*4. Reward Definition*

The local reward $r_i$ at each intersection i is taken as the negative sum of the normalized delay for all vehicles V on all approaches of the intersection as in

$$r_i^t = -\sum_{v \in V} \frac{d_v^t}{d_{max}}. \quad (11)$$

In (11), $d_v^t$ is the individual vehicle delay at time t of each vehicle on the intersection approaches, while $d_{max}$ is the maximum expected vehicle delay for that intersection. By using the ratio $d_v^t/d_{max}$, the delay is scaled between 0 and 1. The delay of each vehicle used in the computation of reward at the intersection excludes the delay encountered at previous intersections. This avoids penalizing the side streets in favor of the main line traffic with accumulated delay from other intersections. As shown by previous studies, the summation of normalized delay in the reward tends to output the direct summation of delay values [7].

## III. Case study

### *A. Test Network*

The selected test network is a section of North Ave corridor in Midtown, Atlanta. Fig. 4 is a google map snapshot showing the test corridor and the surrounding area. The section consists of seven signalized intersections marked with red circles on the figure. Moving from west to east the included intersections are North Ave with Tech Pkwy. NW, Techwood Dr. NW, I-75/I-85 off-ramp, Spring St. NW, W. Peachtree St. NW, Peachtree St. NE, and Jupiter St. NE.

### *B. Data*

The North Ave main street varies between a 4-lane and 6-lane facility, with turn bays at several intersections. Spring St. NW, W. Peachtree NW, and Jupiter St. NE are one-way streets. All the seven intersections run on Qfree Advanced Traffic Controllers (ATC), utilizing the MaxTime® traffic signal control software. For the signals, the Georgia Department of Transportation (GDOT), as part of an ongoing research project, provided detector layout maps and the signal timing databases. The signal timing databases include detector settings, controller settings, and signal timing plans. For this effort the PM peak was utilized. The existing corridor operates actuated-coordinated signal timing plans with a cycle length of 120 seconds in the PM peak. The coordinated movements are the North Ave. westbound and eastbound through movements, except for North Ave. @ Spring St. NW where the Spring St. southbound movement is coordinated.

Fig. 5 shows the PM peak (04:30-05:30 pm) turning movement counts at the seven intersections collected on April, 04th 2023. The data was collected semi-automatically with videos cameras. The volumes in the figure are in vehicles per hour occurring over the peak hour.

Additional trajectory data (i.e., vehicle location and speed) was drawn from drone videos as part of an ongoing GDOT research project. The drone data was collected on April 04th, 2023, in the PM peak from 3 pm to 7 pm. The data was collected with 6 drones to capture different angles and sections of the corridor. Field collected drone video data was processed using DataFromSky Software tools. The trajectory data was then utilized for simulation model calibration, discussed in section C below.

### *C. Model Development and Calibration*

#### *1. Simulation Model Development*

PTV-Vissim® is selected as the simulation environment. PTV-Vissim® is a microscopic, multi-modal traffic simulation software widely used in industry and research.

The network geometry is drawn from google maps and confirmed through site visits. Model construction, quality control, and quality assurance follow the guidance laid out in [35]. Fig. 6 shows the network model in PTV-Vissim®.

For the existing conditions simulation model, the r emulator is utilized to control the intersections, forming the PTV-Vissim-MaxTime® SILs. The field controller MaxTime® databases with the field signal timing plans are loaded directly into the MaxTime® emulator which ensures that the existing conditions simulation runs the same signal timing plans and execution as the field. Each instance of MaxTime® (one for each intersection) is connected to PTV-Vissim through a unique transmission control protocol (TCP) port. A python event-based script embedded in Vissim is used to open and close MaxTime® instances at the beginning and ending of each simulation run, respectively.

The RL-based simulation similarly leverages the PTV-Vissim. However, the phase changes follow the agent's actions. Phase changes are implemented directly in Vissim with signal indication status controlled through COM.

#### *2. Model Calibration*

Model calibration involves adjusting the model parameters so that the model performance matches field conditions [35]. Model calibration was performed utilizing the vehicle trajectories drawn from the drone videos. The annotated trajectories were post processed to extract vehicle speeds, headways, accelerations, etc.

[35] was used as guidance for the PTV-Vissim® model calibration. For this effort two sets of parameters are calibrated, (1) speed distribution and (2) car following model parameters. The desired speed distribution parameters in PTV-Vissim® are adjusted such that that the desired speed distribution matches the field observed speed distribution. For car following, the parameters of Wiedemann 74 flow model, that is, ax, $bx_{Add}$, $bx_{Mult}$, are adjusted such that the PTV-Vissim® simulated saturation headway distribution reasonably match field measured headway distribution. Fig. 7 shows a comparison of cumulative distribution function (CDF) of field-measured headways vs calibrated model headways at North Ave. @ Spring St. The final calibrated model parameters are ax = 9, $bx_{Add}$ = 2.2, and $bx_{Mult}$ = 3.2.

### *D. RL and Simulation Architecture*

Seven agents are formulated, one for each intersection. Fig. 8 shows the interaction of agents with the PTV-Vissim® simulation environment. At each time step, each agent k pulls input from PTV-Vissim® including the local state $S_k$, and the global state $Sc_k$. Using the local state as input, each agent k selects action $a_k$, forming a set of global actions $a_1$ to $a_7$.

Fig. 9 shows the detailed interaction between the agents (1-7) and PTV-Vissim® during the simulation run. To integrate RL into PTV-Vissim® the Component Object Model (COM), an application programming interface (API) for PTV-Vissim®, allows interaction with the simulation to extract data and to read and write to different objects during simulation runtime. This provides the means to extract vehicle data and performance measures to formulate state and rewards functions and to implement the selected actions in form of signal displays. The utilized architecture utilizes event-based scripts embedded directly in the PTV-Vissim® interface rather than external COM functions. This follows from a recent study by [23] that showed that executing PTV-Vissim® with event scripts is many fold faster than using external COM scripts. Three python scripts are embedded in PTV-Vissim®, the “State & Reward Script”, the “Agent Script” and the “Model Runner Script”. The Model Runner Script (MRS) is the central script tracking the simulation time and algorithm time step, initiating actions of the former two scripts and providing signal timings to PTV-Vissim® for implementation.

At a timestep t, MRS passes the current local states ($s_{1\text{-}7}$) to the agents’ actor networks ($s_1$ to agent 1 and so forth) to estimate the logits and actions. At the same time, MRS passes the global states $Sck_{1\text{-}7}$ , one for each agent’s critic network to estimate the state values $V(Sck_{1\text{-}7})$ for that time step. With the selected actions, MRS determines Δt as described earlier. The selected actions for each agent are converted into signal timings and sent to PTV-Vissim® for display for a duration of Δt. The minimum green and yellow and clearance times are set to the same values present in the field signal timing plans. After Δt, MRS requests the next local state, global state and reward for each agent from the State & Reward Script. The State & Reward Script extracts vehicle records and signal states from PTV-Vissim® to formulate local state, global state, and reward functions as discussed in section II.B (8) through (11). A tuple of local state, global states, actions, rewards, logits and state values ($s_{1\text{-}7}$, $Sck_{1\text{-}7}$, $a_{1\text{-}7}$, $logits_{1\text{-}7}$, $r_{1\text{-}7}$, $V(Sck_{1\text{-}7})$ with 1to 7 corresponding to for each agent are saved in the memory buffer for each agent for training at episode end.

*1. Agent Architecture, Training and Hyperparameter Selection*

All the seven agents are formulated with the same architecture for the actor and critic networks. Preliminary training is performed to fine tune the hyperparameters. The finally selected actor network has two hidden layers with 64 and 128 neurons, a clip ratio of 0.2, and a learning rate of 0.0003. The finally selected critic network has three hidden layers with 128, 256 and 256 neurons, a learning rate of 0.001 and a discount factor of 0.99.

Fig. 10 shows the training architecture for each agent k. Training is performed at the end of each episode using data archived only that episode (on-policy learning). For a full trajectory over the entire episode using archived rewards and state values, advantage function values are estimated following (1). Using the archived actions, logits, actions and local states, the actor network estimates the importance sampling weights as per (5). Using importance sampling weights and estimated advantage values the actor clipped loss is determined per (6). With the computed loss values, the parameters of the policy/actor network are updated using gradient accent. Using the archived global state values and rewards, the critic loss is computed as per (7). The parameters of the critic network are then update via gradient descent.

*2. MA-PPO Testing*

Fig. 11 shows the algorithm architecture during testing. As indicated, critic networks are only active during training but not during implementation/testing. During testing, agents 1-7, take local states ($s_{1\text{-}7}$) as input to predict the actions/signal timings that are implemented locally at each intersection. For each of the tests described in the section E, ten replicate simulation runs are performed, each with a unique random seed. Each simulation run lasts for one hour with data collected only in the last 45 minutes, with the first 15 minutes of the hour taken as the warm-up period.

*E. Results*

*1. Model Training*

Fig. 12 shows the learning curves for the seven agents, one for each intersection. The x-axis plots the episode number while the y-axis plots the reward. Each data point is the average of rewards for all steps within the episode. Each episode lasts for 30 simulation minutes. Simulation runs and model training are performed on an x64-based PC equipped with 12th Gen Intel(R) Core i9-12900, 2400 MHZ, 16 Core(s), 24 Logical Processor(s),128 GB of RAM, Intel (R) UHD Graphics 770 GPU and Windows 11 operating system. It required approximately 60 hours to reach a fully trained model with the main limiting factor being PTV-Vissim®’s runtime efficiency as fully discussed in a recent study by [23].

The first four agents listed (North Ave. at Jupiter St., Peachtree St. NE, W. Peachtree St., and Spring St. NW) converge within the first 300 episodes. North Ave at I-75/I-85 off-ramp also converges within the first 300 episodes except for continued variability which eventually dies down. These intersections each have a maximum of four phases and despite the higher or comparable traffic volumes at these intersections, convergence occurs significantly faster than at intersections with additional phases.

North Ave. at Tech Pkwy. NW (7 phases) converges after approximately 600 episodes while North Ave. at Techwood Dr. NW (8 phases) requires more than 1000 episodes to converge. The complexity of signal plans, including the number and sequence of phases appears to be the primary factor affecting the convergence. This is intuitively reasonable as the agents have larger action spaces.

*2. MA-PPO vs field implemented coordinated ASC*

The performance of the trained MA-PPO is evaluated against the field implemented signal timing plans using delay and travel time as performance measures. Fig. 13 shows a comparison of main line movement travel time from the trained MA-PPO adaptive signal control and the field implemented coordinated ASC plans modeled using the MaxTime® emulator. The test volumes are the PM peak field volumes as shown in Fig. 5. The two movements are defined on the main street, from end to end of the model, in the eastbound (EB) and westbound (WB) direction. EB travel time is measured from a point upstream of North Ave @ Techpkwy to a point downstream of North Ave @Jupiter St NW while WB travel time is measured from a point upstream of North Ave @Jupiter St NW to a point downstream of North Ave @ Techpkwy. Each box plot is formed from ten data points, the average travel time for all vehicles in each of the ten replicate simulation runs.

In the WB which is the primary coordination direction for ASC, both MA-PPO and ASC show very comparable performance, with MA-PPO showing slightly lower travel time. However, in the WB, MA-PPO performed significantly better than ASC. Compared to the field implemented plans, MA-PPO showed a travel time improvement of 24% in the WB direction. Coordinated ASC is typically optimized to favor progression in a primary coordination direction, with the secondary direction likely receiving poorer service. This approach may be sub-optimal, especially if there is a high proportion of traffic in the secondary coordination direction. This is the case on North Ave. corridor. The superiority of MA-PPO comes from the ability to optimize performance in both directions.

Fig. 14 shows the performance of the two control systems for the cross street through movements using delay as the performance measure. MA_PPO shows significantly less delay for all the measured cross street movements (only through movements shown here for brevity). This includes Spring St. southbound which is the coordinated movement in the field ASC signal timings. Thus, the selected reward system and the centralized training leads to better performance for both main and cross streets at all intersections compared to the field implemented ASC.

### *3. Volume Sensitivity and MA-PPO Robustness*

The above result is based on a single day field volume set and quite possibly may not reflect the typical day for which the ASC signal timings are optimized. To capture the impact of variability of traffic volumes and further test the robustness of the trained MA-PPO, sensitivity experiments are performed for two other volumes sets. Keeping the origin-destination (O-D) ratios fixed, the model entry volumes are adjusted (a) higher by 5% and (b) lower by 10% from the field measured volumes of Fig. 5. The limit of 5% higher is considered as some intersections are at or close to capacity for the field volumes. Fig. 15 shows the main street travel time from MA_PPO and ASC for the three volume levels. Travel time is measured from end to end of the model as described in section 2 above2. As expected, travel time increases as the volumes increase for both control systems. MA-PPO consistently outperforms ASC at all the three volume levels.

Fig. 16 shows a similar comparison for the cross streets, using delay as the performance measure. As with the North Ave. mainline, MA-PPO consistently outperforms ASC at the three volume levels. This demonstrates the robustness of the trained model to adapt to variations in traffic demand.

## IV. CONCLUSION

This study formulates a multi-agent proximal policy optimization algorithm to implement adaptive and coordinated traffic control along an arterial corridor. The formulated MA-PPO has centralized critic architecture in which each agent has an actor network that selects independent actions conditioning on local observations and a centralized critic that estimates the agent's value function conditioning on global observations. All agents are formulated to allow selection and implementation of up to eight signal phases as commonly implemented in the field controllers. The formulated algorithm is tested on a simulated real-world corridor with seven intersections. The performance of the formulated MA-PPO adaptive control algorithm is compared with the field implemented ASC signal timing plans modeled using PTV-Vissim®-MaxTime® SILs.

The speed of convergence for each agent largely depended on the size of the action space which in turn depended on the number and sequence of signal phases. The intersections with 4 or fewer phases converged within 300 episodes each of 30 simulation minutes while the intersections with 7 and 8 phases required double to triple the number of episodes for convergence.

For the field measured traffic volumes, for all movements, the trained MA-PPO performed significantly better than the field implemented ASC signal timings. For the main street, MA-PPO optimizes performance in two directions compared to ASC which largely optimizes performance in the primary coordinated direction. Using travel time as the performance measure, the two control systems showed very comparable performance in the WB direction which is the PM peak direction and the primary coordination direction for ASC. In the EB direction, MA-PPO showed a 24% reduction in travel time compared to ASC. For cross streets, MA-PPO showed significantly reduced delay

To capture the variability of traffic volumes and further test the robustness of the trained MA-PPO, sensitivity experiments are performed for two other volumes sets, (a) 5% above and (b) 10% below the field measured volumes. For both main street movements and cross streets movements, MA-PPO consistently outperformed ASC at the three volume levels. This demonstrated the robustness of the trained model to adapt to variations in traffic demand.

A key challenge of training of RL agents in a high-fidelity traffic microsimulation environment like VISSIM® is run time efficiency to reach hundreds to thousands of simulations runs required to fully train the agents. This study uses online serialized training for the formulated agents. Sixty hours were required to complete training of the agents. This limited the exploration and experimentation that could be achieved. Ongoing follow up studies are considering parallelized training architectures to reduce the required simulation run time.

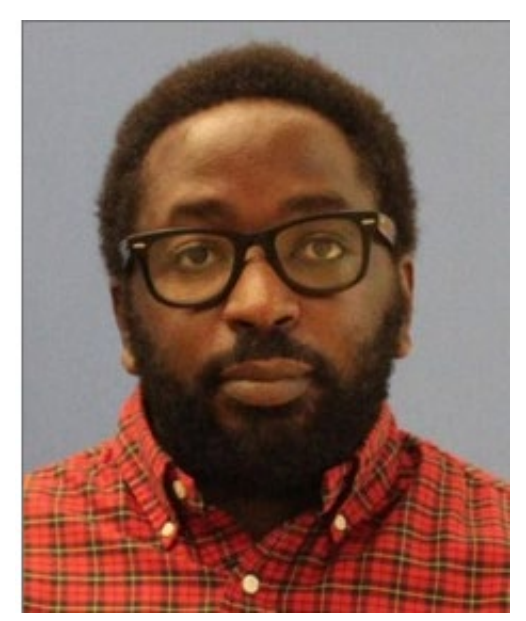

**DICKENS KAKITAHI KWESIGA** is a PhD student and a graduate research assistant in the School of Civil Engineering at Georgia Institute of Technology. His research interests include modeling, simulation and optimization of arterial corridor operations, traffic signal systems, Transit signal priority, emergency vehicle preemption, connected vehicles and AI based traffic control and AI/ML applications in traffic operations. His email is dkwesiga3@gatech.edu

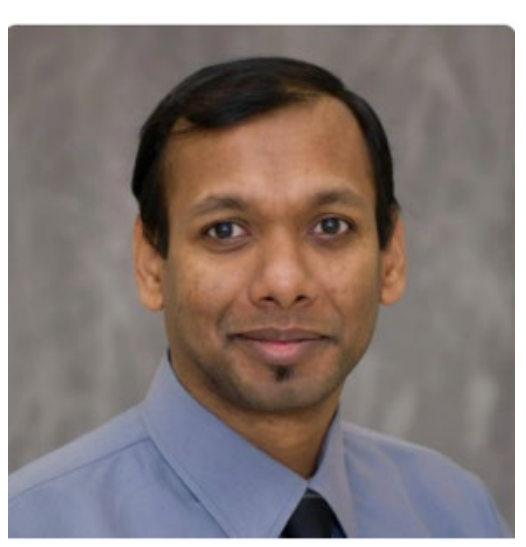

**ANGSHUMAN GUIN** is a Senior Research Engineer in the School of Civil and Environmental Engineering at the Georgia Institute of Technology. His research attempts to find answers through innovations in the development of effective means of data collection, quality assurance, and processing to convert these data into informative metrics across a range of time-scales from near real time to decades. Dr. Guin's current research projects at Georgia Tech are broadly in Freeway Operations, Connected and Autonomous Vehicles, Intelligent Transportation Systems (ITS), Transportation Safety, Traffic Simulation and Data Management. His email is angshuman.guin@ce.gatech.edu

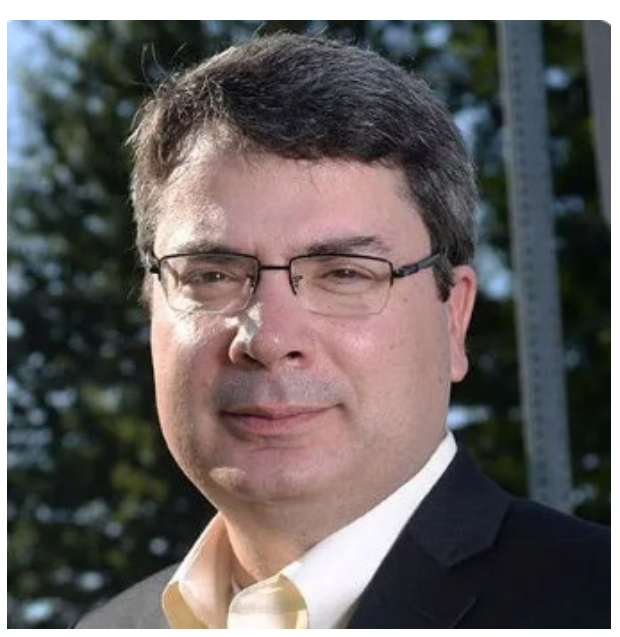

**MICHAEL HUNTER** is a Professor in the School of Civil and Environmental Engineering at Georgia Institute of Technology. His primary teaching and research interests are in transportation operations and design, specializing in emerging technology, adaptive signal control, simulation, and arterial corridor operations. His email is michael.hunter@ce.gatech.edu

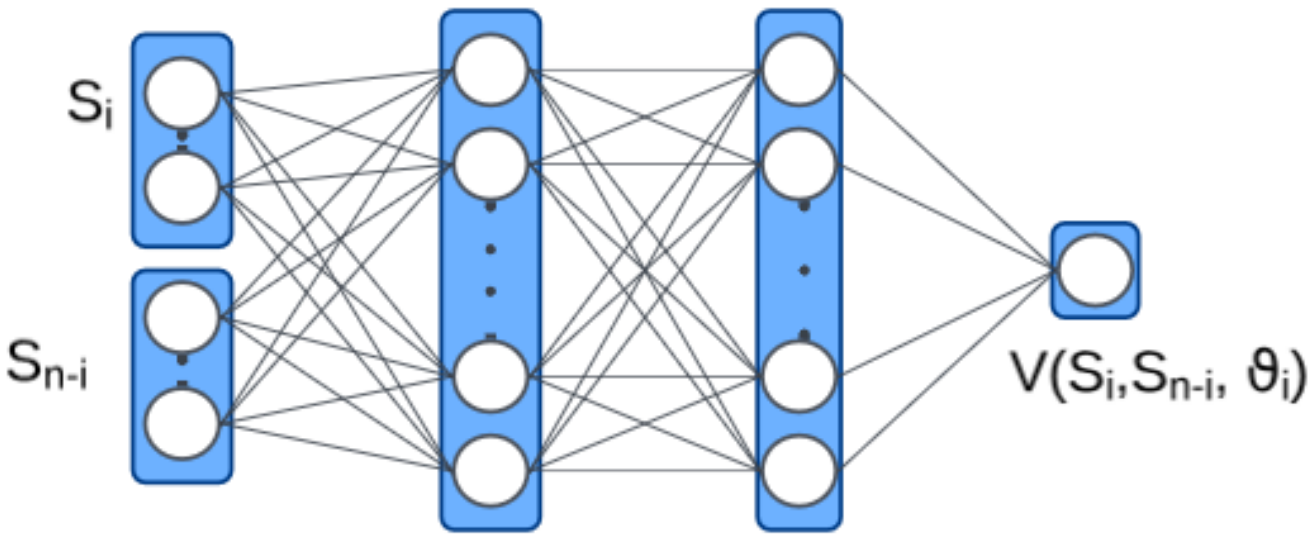

Fig. 1. Architecture of Centralized critic for agent i, adapted from [1]

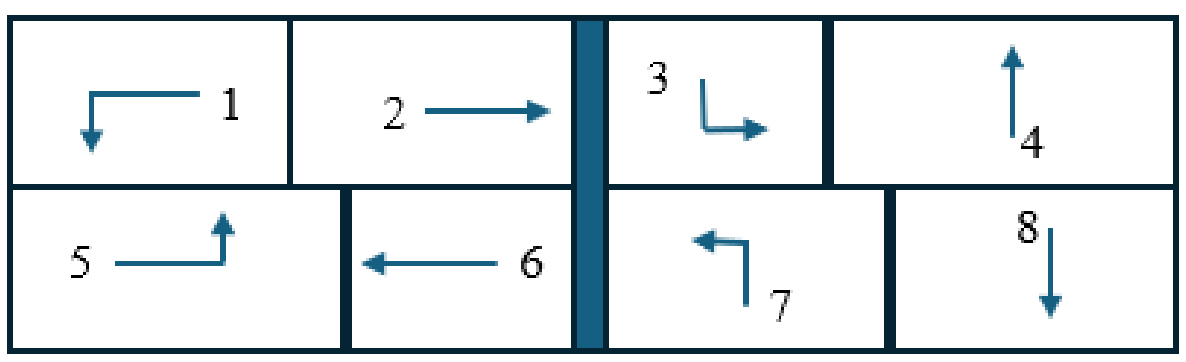

Fig. 2. Dual ring barrier diagram

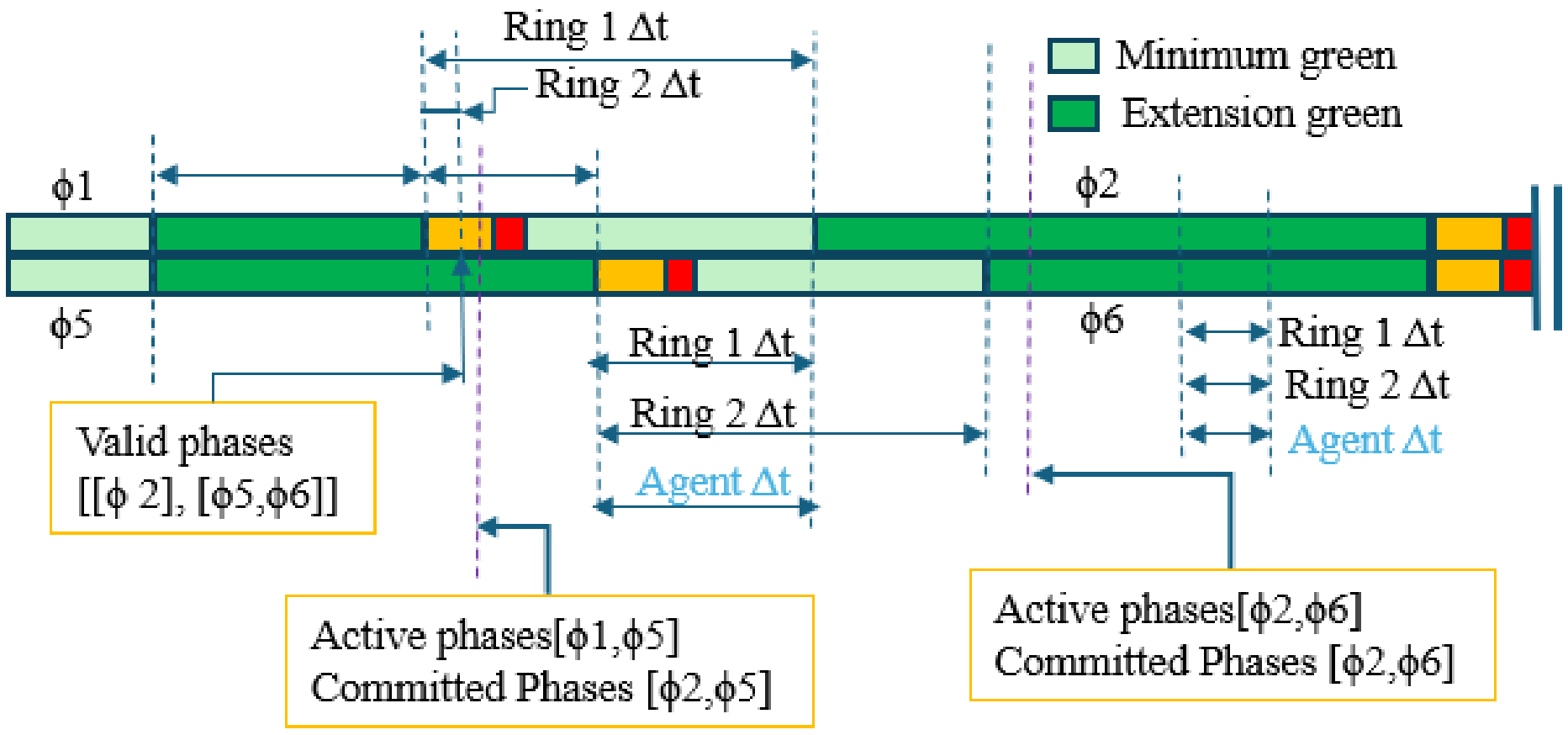

Fig. 3. Action Implementation in dual ring barrier configuration

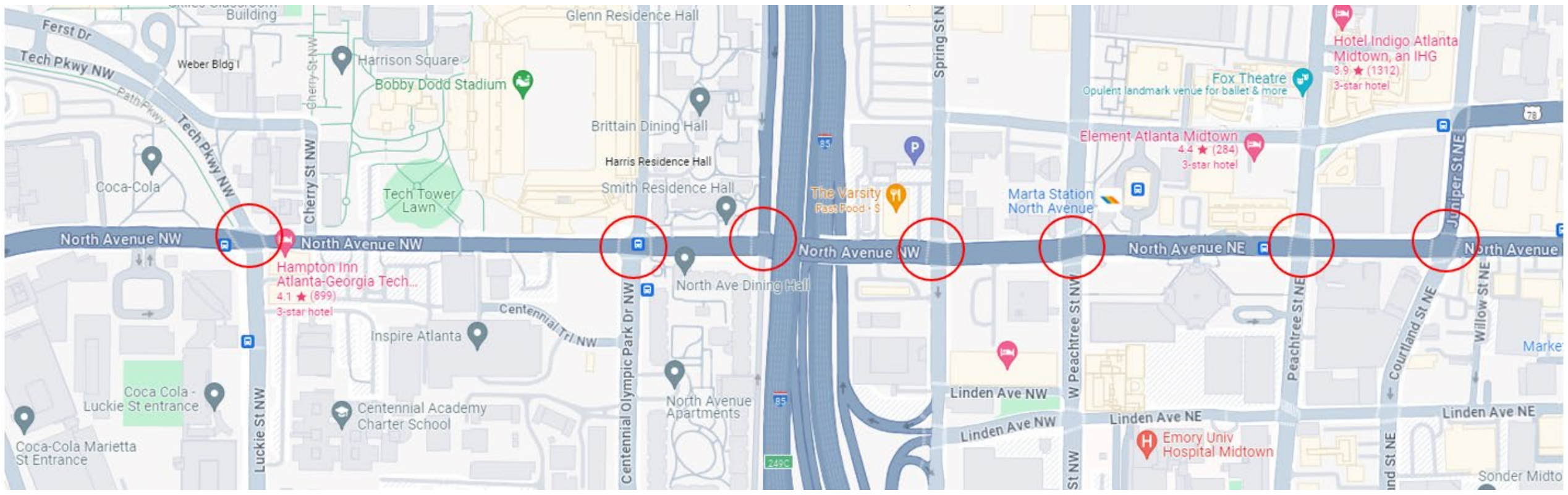

Fig. 4. Test network extents

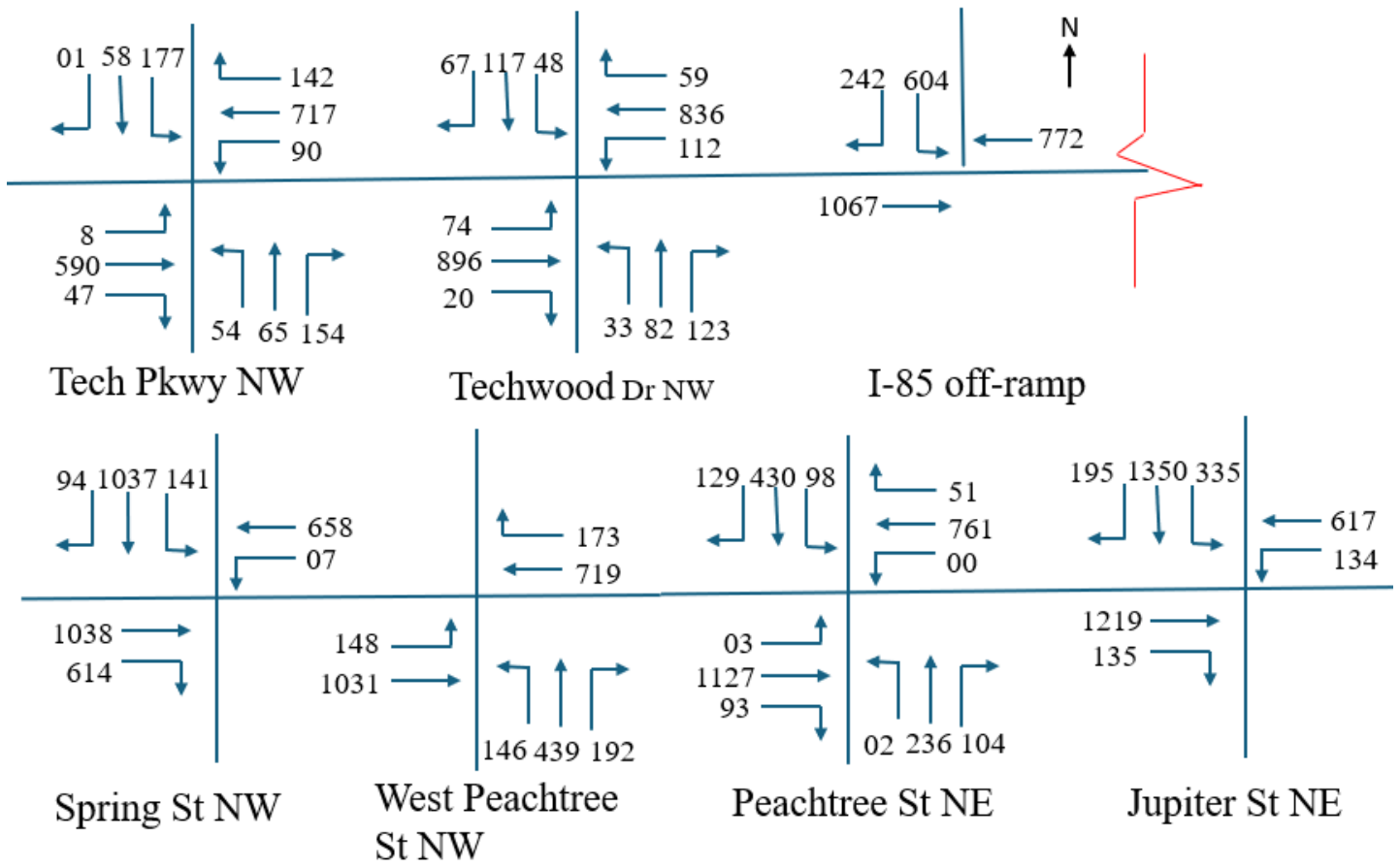

Fig. 5. PM peak volumes

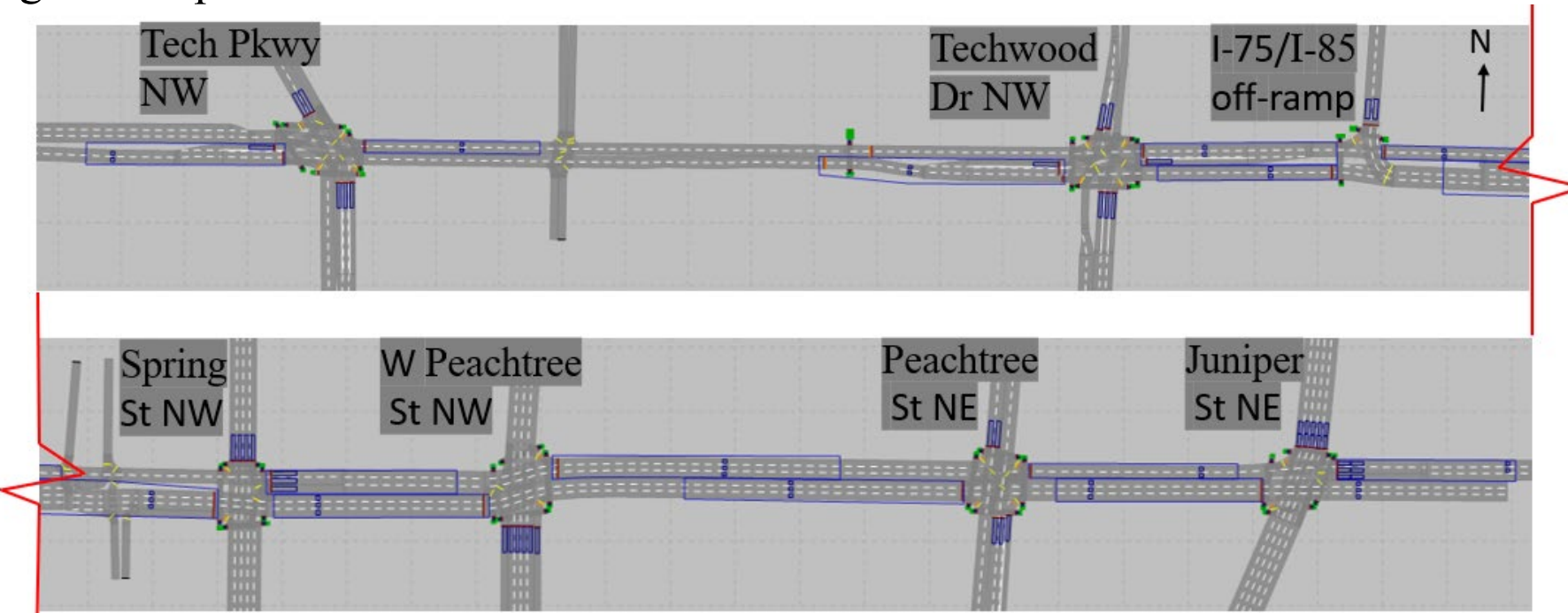

Fig. 6. Network Model in PTV-Vissim®

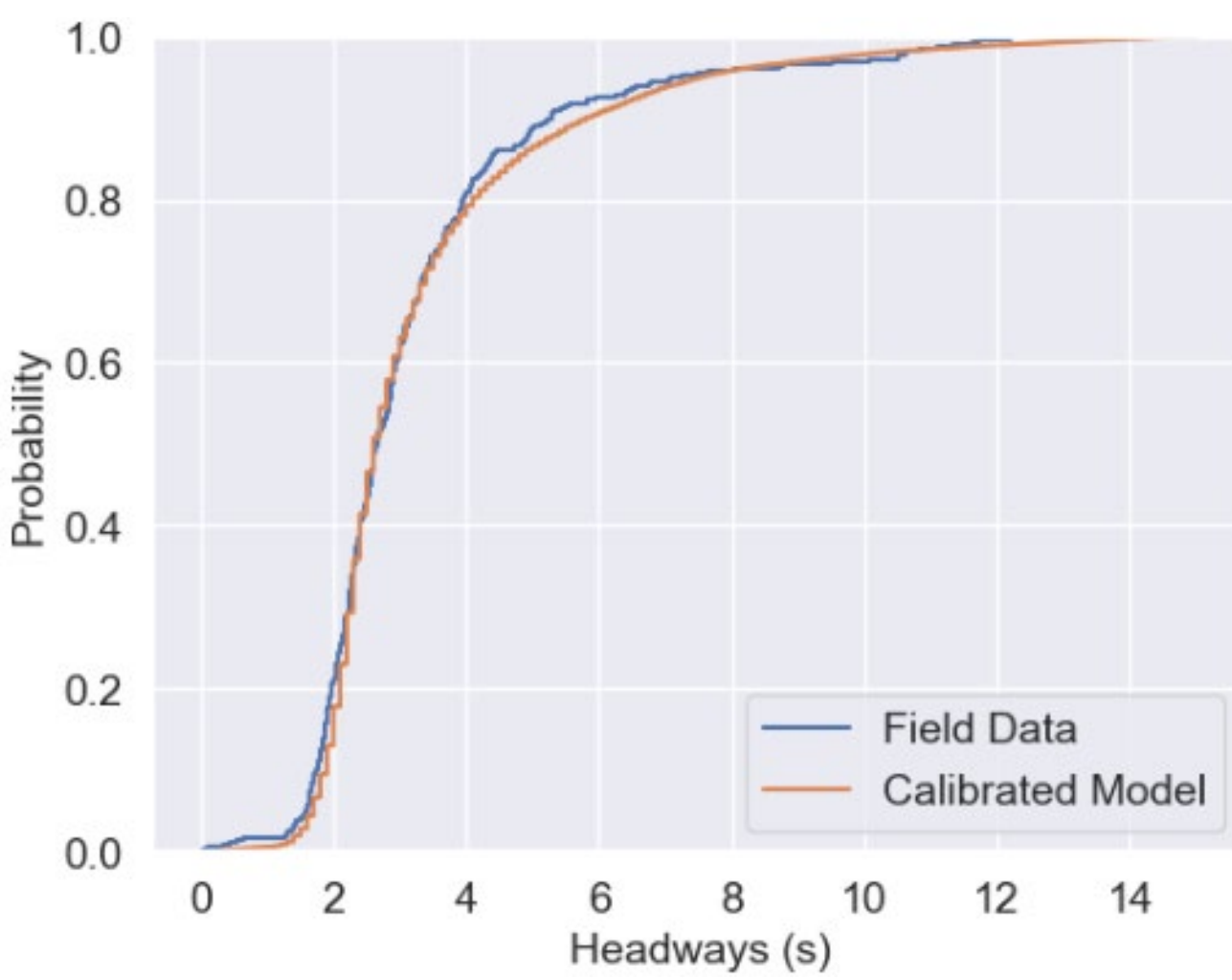

Fig. 7. Field Vs. calibrated model headways at North Ave. @ Spring St.

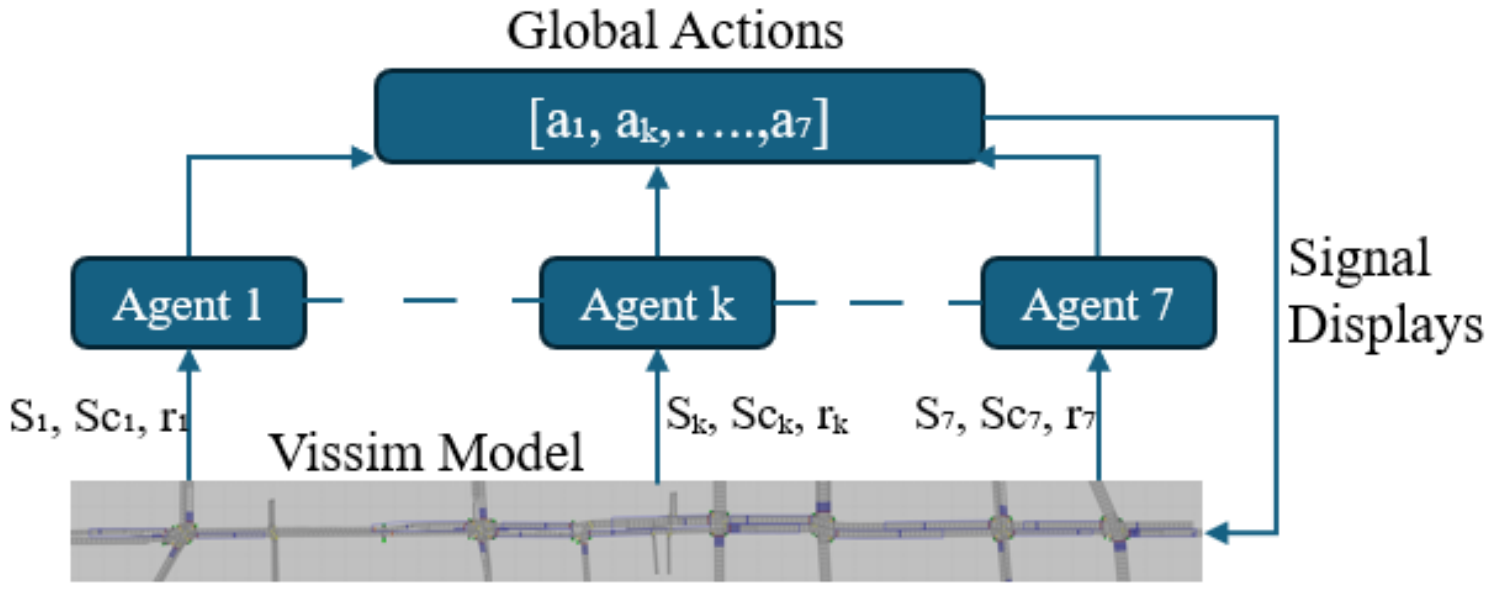

Fig. 8. Interaction of agents with PTV-Vissim® Simulation environment

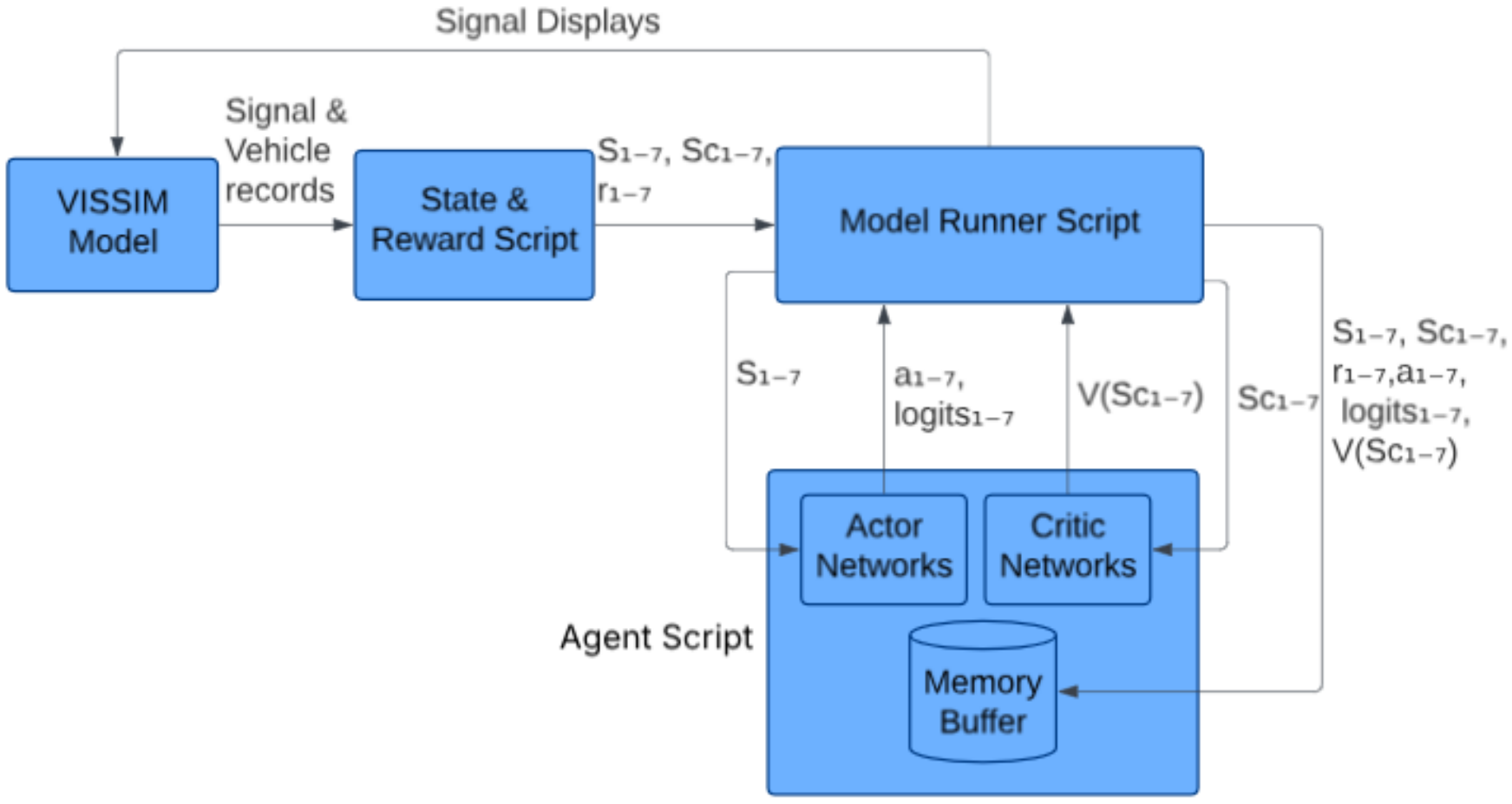

Fig. 9. Agent-Vissim interaction during simulation run

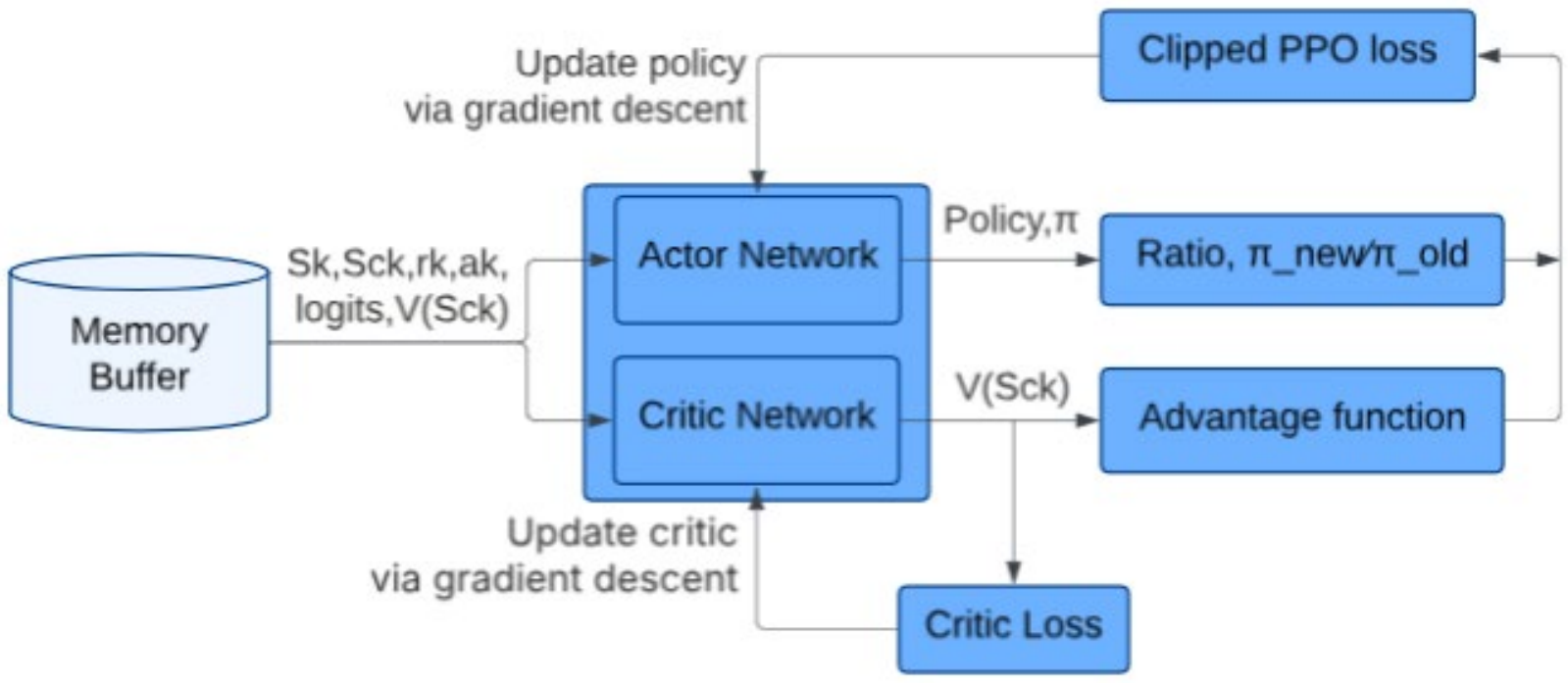

Fig. 10. Algorithm architecture during training

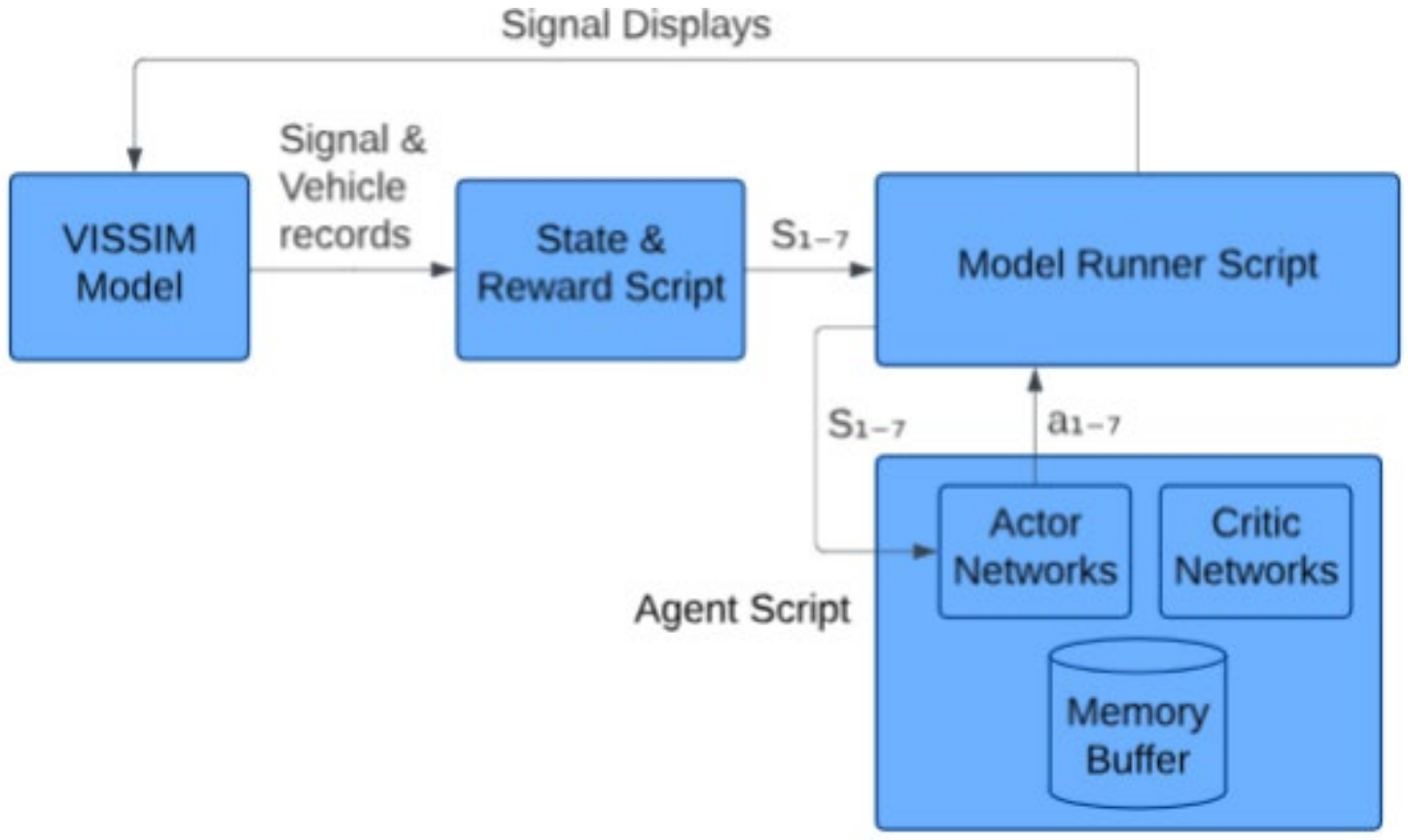

Fig. 11. Algorithm architecture during testing

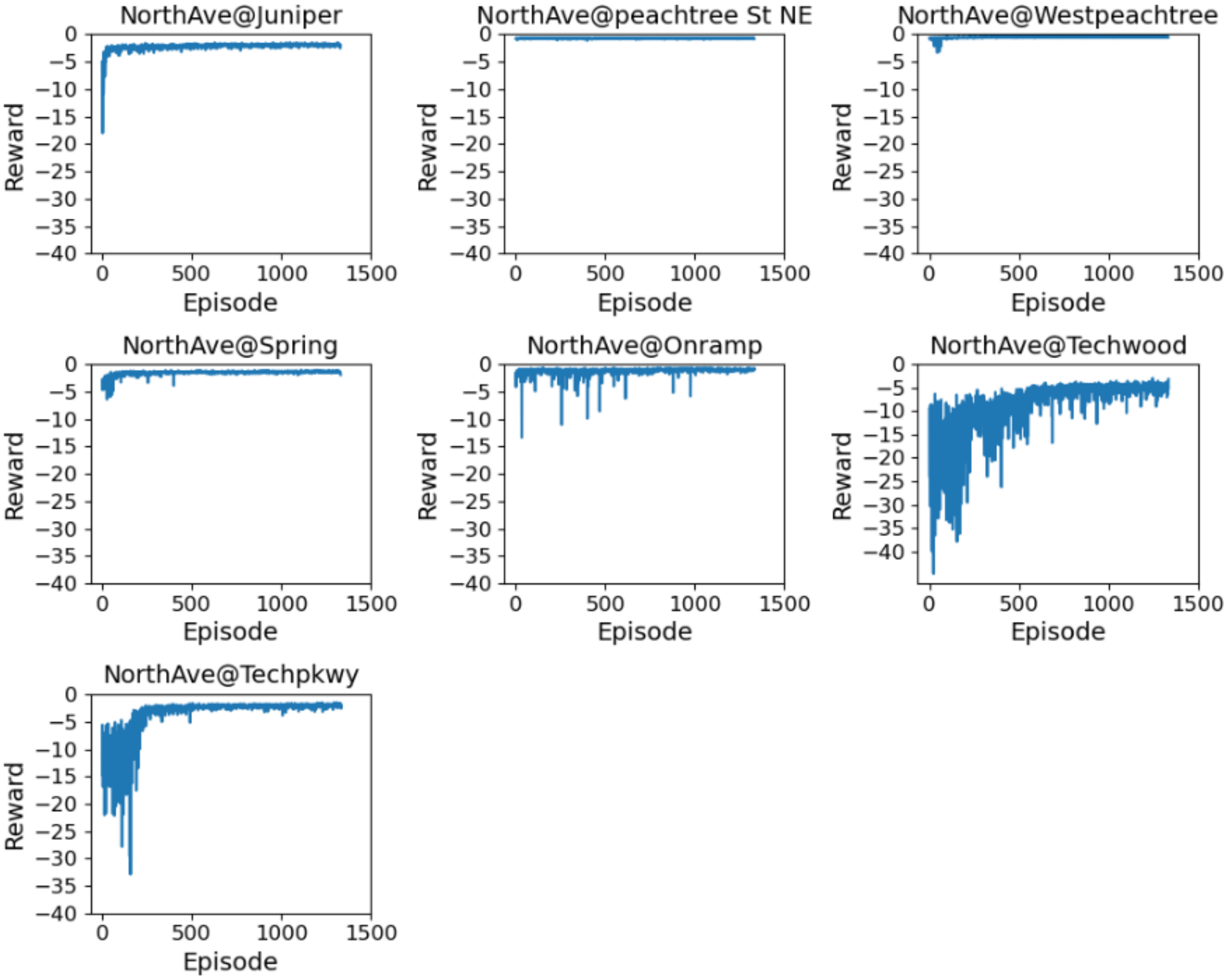

Fig. 12. Learning Curves for different intersection agents

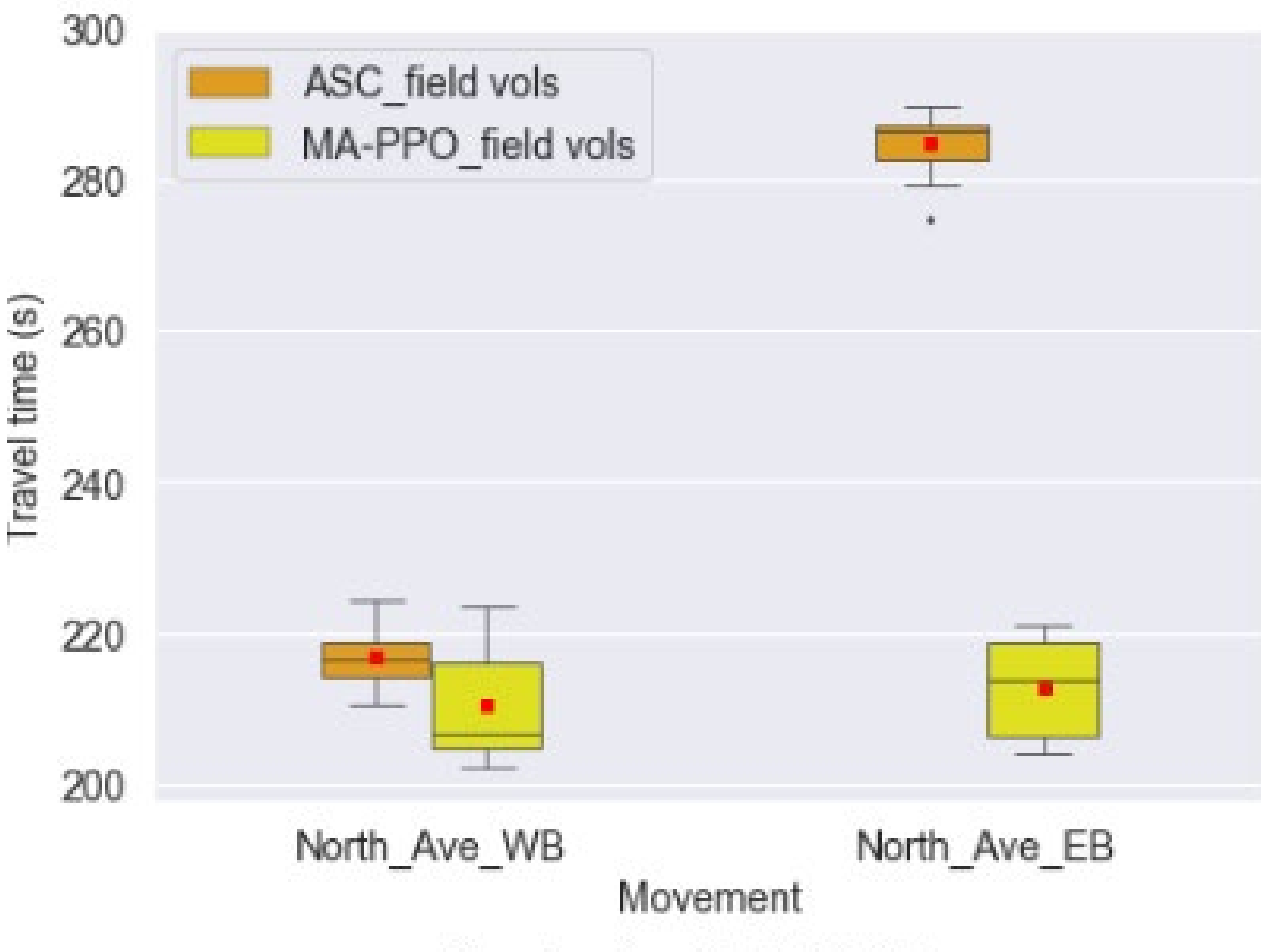

Fig. 13. Main line movement travel time for ASC and MA-PPO for field measured traffic volumes

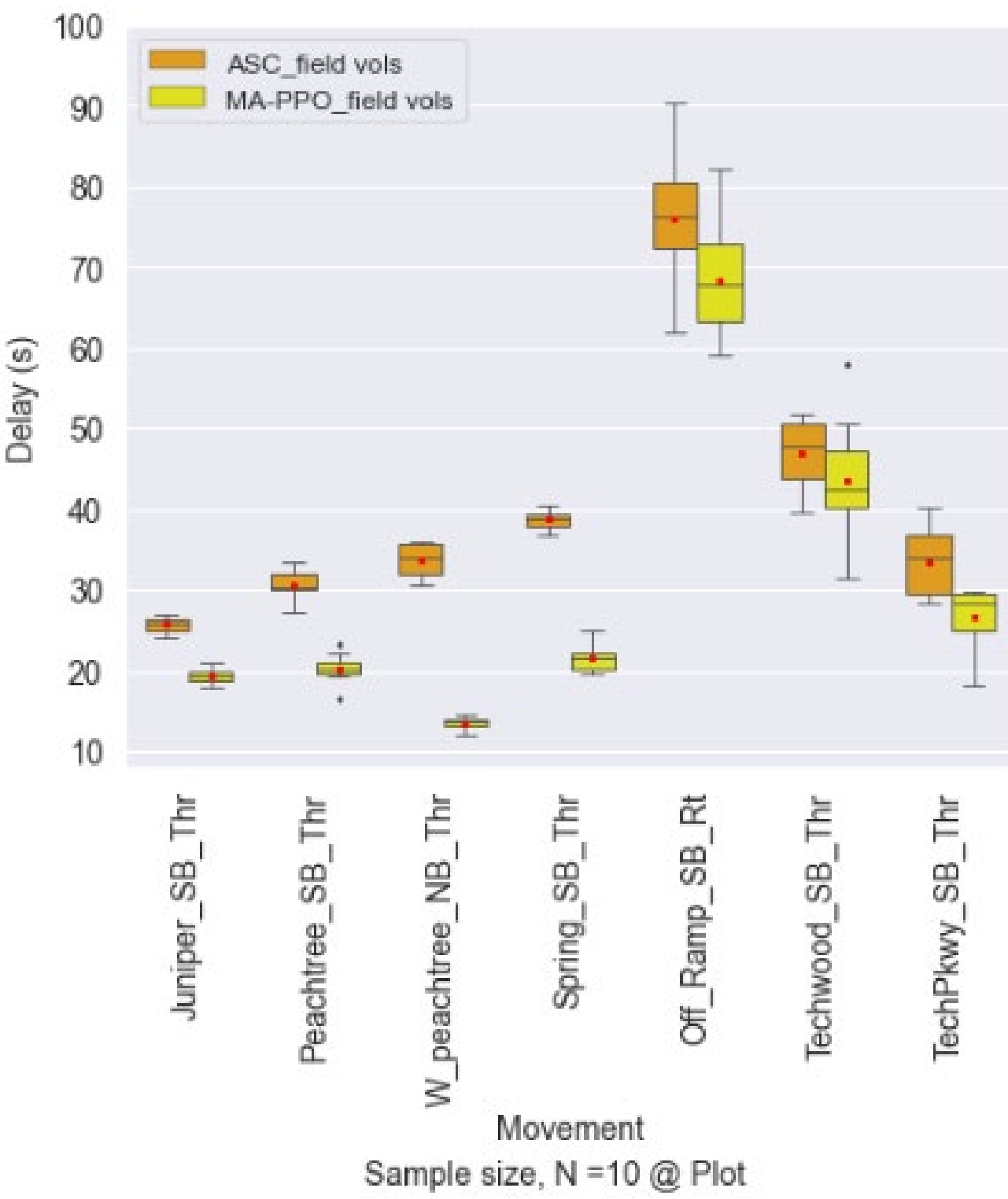

Fig. 14. Cross street through movement Delays for MaxTime® and MA-PPO for field measured traffic volumes

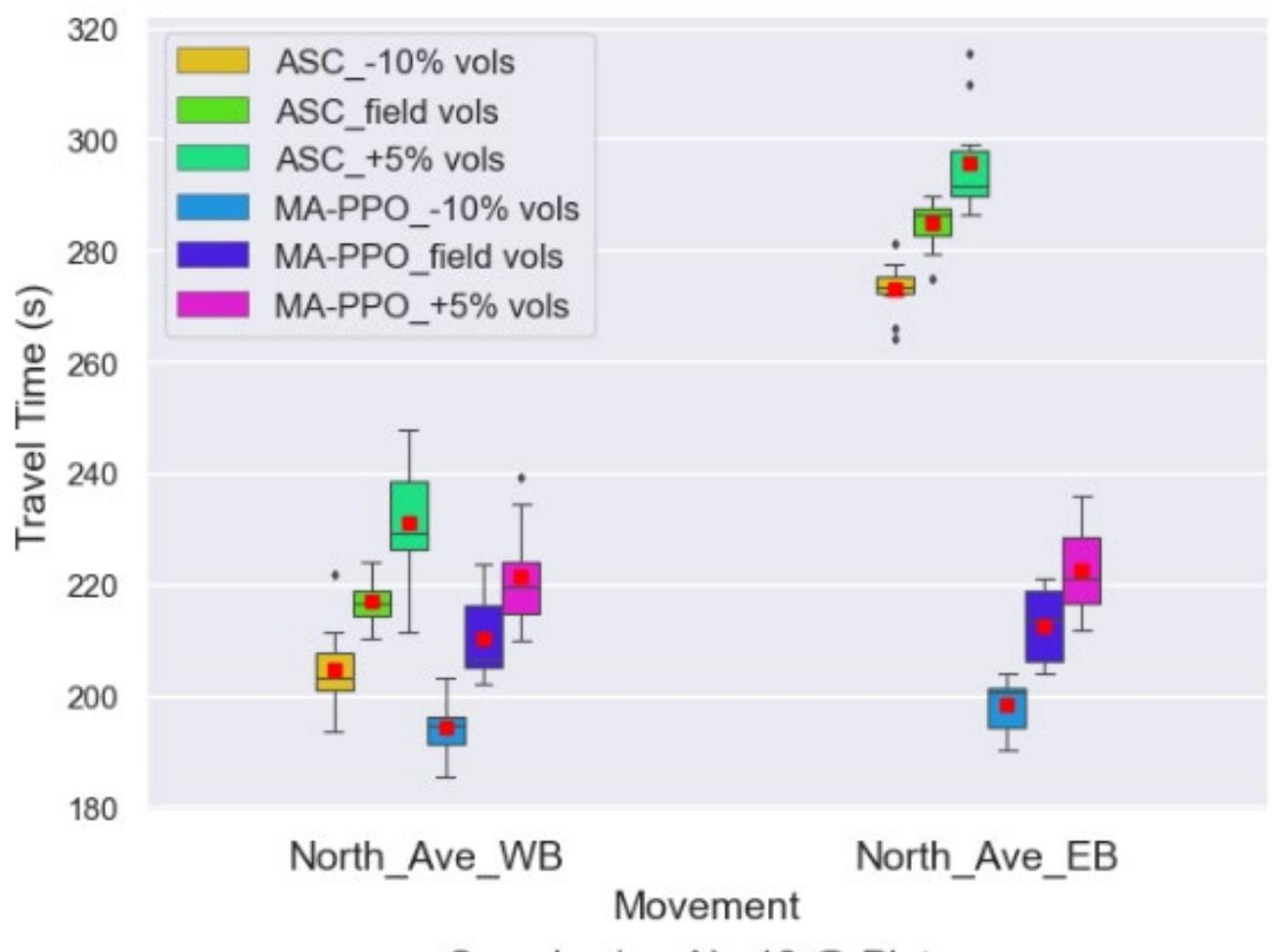

Fig. 15. Main street travel time from MA-PPO vs coordinated ASC at three volume levels

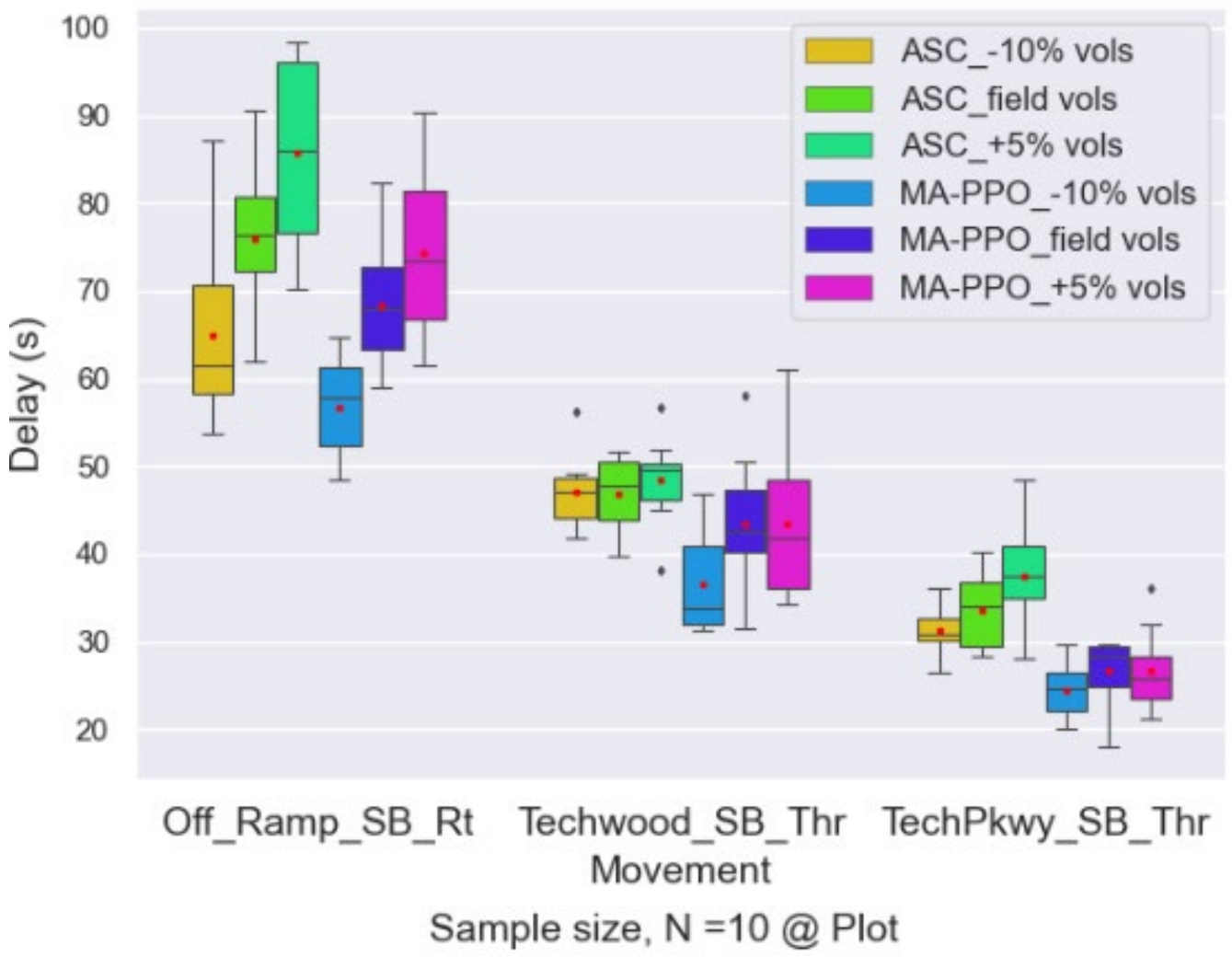

Fig. 16. Cross street Delay from MA-PPO vs coordinated ASC at three volume levels